# Inverse-Designed Chiral Metasurfaces for Enhanced Helical Dichroism

Chia-Chun Pan[1], Yu-Hsi Chen[2], Munseong Bae[3], Haejun Chung[3], Dan Smith[4], Daniel Fan[1], Sejeong Kim[5,6,*], Ranjith R Unnithan[1,*]

[1]Department of Electrical and Electronic Engineering, University of Melbourne, Melbourne, VIC, Australia

[2]School of Mathematics and Statistics, University of Melbourne, Melbourne, VIC, Australia

[3]Department of Electronic Engineering, Hanyang University, Seoul 04763, South Korea

[4]Materials Characterisation and Fabrication Platform, University of Melbourne, Melbourne, VIC, Australia

[5]Department of Electrical and Computer Engineering, Sungkyunkwan University (SKKU), Suwon 16419, Republic of Korea

[6]Center for 2D Quantum Heterostructures (2DQH), Institute for Basic Science (IBS), Sungkyunkwan University (SKKU), Suwon 16419, Republic of Korea

Corresponding authors: sejeongk@skku.edu, r.ranjith@unimelb.edu.au

## KEYWORDS

inverse design, orbital angular momentum, helical dichroism, adjoint method, chiral metasurfaces, nanophotonics

## ABSTRACT

Helical dichroism (HD), arising from the interaction between structured light carrying orbital angular momentum (OAM) and chiral matter, offers opportunities for enhanced chiroptical characterization at the subwavelength scale. Here, we employ inverse design to develop chiral metasurfaces tailored for enhanced HD under Laguerre-Gaussian (LG) beam illumination. The optimized planar structure exhibits a simulated HD around 115% at 800 nm for opposite OAM states, revealing the strong chiroptical response enabled by the inverse-designed geometry. The optimized metasurfaces were fabricated from $Si_3N_4$ and experimentally characterized using a custom-built optical system with controlled OAM excitation. A single chiral metasurface exhibited a maximum experimental HD up to 62% near 800 nm, demonstrating a strong OAM-dependent helical dichroic response from a single 6-micron-large chiral structure. Although the measured HD value is lower than the optimized simulation value, the experimental result validates the feasibility of inverse-designed chiral metasurfaces for structured-light chiroptics. These findings establish inverse design as an effective approach for engineering compact chiral nanophotonic structures with enhanced interactions with structured light, offering potential for miniaturized chiroptical sensing and OAM-resolved imaging.

**INTRODUCTION**

Chirality underpins a broad range of phenomena in chemistry, biology, and photonics, making its optical characterization central to applications ranging from molecular analysis to nanophotonic sensing. Circular dichroism (CD), characterized by the differential optical response to left- and right-circularly polarized light, provides a direct optical probe of chiral matter by revealing its interaction with the spin angular momentum (SAM) of light [1]. At the molecular and nanoscale levels, however, chiroptical signals are often intrinsically weak, motivating the use of structured optical fields and engineered chiral nanostructures to enhance light-matter interactions [2]. Optical vortex beams introduce an additional degree of freedom through orbital angular momentum (OAM), characterized by the azimuthal phase factor $\exp(-i\ell\phi)$, where $\ell$ denotes the topological charge. The differential response to vortex beams with opposite OAM handedness is commonly called helical dichroism (HD). Recent theoretical and experimental studies have shown that the spatial phase structure and field gradients of OAM beams can induce chiroptical responses governed by both the magnitude and sign of $\ell$, while tight focusing can further convert OAM handedness into local optical chirality, enabling selective coupling to chiral resonances [3, 4].

Experimental studies have established OAM-dependent chiroptical responses across a range of length scales and structural platforms. Early demonstrations showed giant vortical differential scattering from femtosecond-laser-fabricated chiral microstructures, highlighting the critical role of dimensional matching between the incident vortex field and the structure [5]. This concept was subsequently extended to the nanoscale, where planar-chiral nanostructures exhibited HD of about 20% at visible wavelengths [6]. Three-dimensional chiral microstructures and oligomers further demonstrated HD responses of around 23% and 50%, respectively, highlighting the potential of engineered chiral architectures for enhancing OAM-dependent optical responses [7, 8]. More recent studies have further explored HD through structured illumination, resonant excitation, dynamic reconfiguration, and chiral molecular assemblies, demonstrating the versatility of OAM-dependent chiroptical interactions across different material platforms [9-12]. Importantly, these studies also highlight the role of spatial and modal matching between the structured illumination and the chiral structure in determining the resulting HD response [13, 14].

However, most existing approaches rely on manually designed geometries or a limited number of reconfigurable parameters, leaving the high-dimensional design space of chiral structures largely unexplored for systematic and automated optimization of HD. Therefore, on the computational side, adjoint-based topology optimization has been adopted to explore this design space, with numerical studies achieving an HD response of roughly 107% at 800 nm in an inverse-designed $Si_3N_4$ structure under OAM states with $|\ell| = 3$, while enforcing minimum feature-size and gap constraints for fabrication [15]. Another numerical study based on parameterized resonant geometries has also achieved an HD of up to 108% in a deformed hexagonal silicon nanostructure [16]. Hence, we employ adjoint-based topology optimization to design a planar $Si_3N_4$ chiral nanostructure under fabrication constraints and experimentally validate its OAM-dependent optical response. The structure is optimized for differential reflection of Laguerre-Gaussian beams with $\ell = \pm 2$ at 800 nm, yielding a simulated HD of around 115%. Using a custom-built optical setup with controlled OAM excitation, we observe a pronounced difference in the reflected spectra for $\ell = -2$ and $\ell = +2$, corresponding to an experimental HD of approximately 62%. These results demonstrate an OAM-dependent differential response enabled by inverse design in a fabricated dielectric nanostructure, providing a practical pathway toward compact chiral photonic devices with tailored HD.

## RESULTS AND DISCUSSION

### Inverse Design of Chiral Metasurfaces

The objective of the optimization framework was to achieve a strong helicity-dependent optical response by maximizing the HD of the dielectric metasurface. Silicon nitride ($Si_3N_4$, n=2.024) was selected because of its negligible absorption and relatively high refractive index in the near-infrared regime, together with its compatibility with established CMOS fabrication processes [17]. Three-dimensional (3D) finite-difference time-domain (FDTD) simulations were performed using the open-source MEEP software package [18] at a wavelength of 800 nm, with a spatial resolution of 20 pixels/$\mu$m. The computational cell spans $9.2 \times 9.2 \times 6.6\,\mu\mathrm{m}^3$ and is terminated by $0.8\,\mu$m-thick perfectly matched layers (PMLs) along all boundaries to absorb outgoing waves and suppress spurious reflections. Within this domain, the active design region occupies a $6 \times 6 \times 1.2\,\mu\mathrm{m}^3$ volume, with a fixed pattern height $1.2\,\mu$m, a transverse $0.8\,\mu$m air buffer (n=1) and a semi-infinite silicon dioxide ($SiO_2$, n=1.45) substrate. To ensure fabrication feasibility, minimum feature-size and minimum gap-size constraints of 0.2 $\mu$m were imposed throughout the optimization.

To probe the chiroptical response of the structure, the excitation source was defined as a linearly polarized optical vortex propagating along the z-axis at an operating wavelength of 800 nm. Specifically, an LG beam was employed, which carries OAM through its azimuthal phase profile. The spatial electric-field distribution of the incident LG mode is expressed as [19]:

$$u_{\ell p}(r,\phi,z') = \frac{C}{w(z')}\left[\frac{r\sqrt{2}}{w(z')}\right]^{|\ell|} L_p^{|\ell|}\left(\frac{2r^2}{w^2(z')}\right)\exp\left(\frac{-r^2}{w^2(z')}\right)\exp(-i\ell\phi)$$

$$\times \exp\left(\frac{ikr^2z'}{2\left(z'^2+z_R^2\right)}\right)\exp(-i\psi(z'))\hat{x} \quad (1)$$

where $C$ is a normalization factor, $L_p^{|\ell|}$ denotes the associated Laguerre polynomial, $p = 0$ is the radial index, and $\ell$ is the topological charge. The transverse spatial coordinates are the radial distance $r = \sqrt{x^2 + y^2}$ and the azimuthal angle $\phi$. $w(z')$ is the beam width at $z'$, $k = 2\pi/\lambda$ is the wavevector, $z_R = \pi w_0^2/\lambda$ is the Rayleigh range, and $\psi(z') = (|\ell| + 1)\arctan(z'/z_R)$ is the Gouy phase, with the beam waist set to $w_0 = 1.13$ $\mu$m. The unit vector $\hat{x}$ indicates the beam is linearly polarized along the x-axis. In contrast to circular polarization, which provides two SAM states only, OAM beams offer a set of orthogonal states characterized by their topological charge $\ell$. This additional degree of freedom enables selective excitation and characterization of chiral light-matter interactions. In this study, the target OAM modes were chosen as $\ell = \pm 2$, as preliminary sweep over different OAM orders confirmed that $\ell = \pm 2$ yielded the highest HD among the tested configurations and was therefore selected for subsequent optimization and experiments. To quantify the chiroptical response, the modal reflectance $R_\ell$ for each incident OAM mode was calculated by integrating the time-averaged Poynting flux over the output monitor surface **S** and normalizing it to the incident source power [15]:

$$R_{\ell=\pm 2} = \left| \frac{\int \mathrm{Re}(\mathbf{E}_r \times \mathbf{H}_r^*) \cdot d\mathbf{S}}{\int \mathrm{Re}(\mathbf{E}_i \times \mathbf{H}_i^*) \cdot d\mathbf{S}} \right|_{\ell=\pm 2} \quad (2)$$

where $\mathbf{E}_i$ and $\mathbf{H}_i$ denote the incident electric and magnetic fields at the reflection monitor and are used for power normalization, while $\mathbf{E}_r$ and $\mathbf{H}_r$ denote the corresponding reflected fields. The surface $\mathbf{S}$ represents the reflection-monitor area over which the Poynting flux is integrated. Building upon these reflection efficiencies, the objective function, or the figure of merit (FoM), was designed to maximize the asymmetric response between the two opposing optical vortices. Specifically, optimization seeks to maximize the reflection efficiency of the target negative topological charge ($R_{\ell=-2}$) while simultaneously suppressing that of the positive charge ($R_{\ell=+2}$). This asymmetric response is quantified by HD, defined as the contrast between the two states:

$$\mathrm{HD} = 2 \times \frac{R_{\ell=-2} - R_{\ell=+2}}{R_{\ell=-2} + R_{\ell=+2}} \times 100\% \quad (3)$$

By optimizing this metric, the inverse-design process iteratively adjusts the material distribution to maximize the HD and thereby achieves a strong asymmetric response.

To navigate the non-intuitive photonic design space, topology optimization is combined with adjoint sensitivity analysis, enabling efficient evaluation of the design gradients through the Born approximation and the Lorentz reciprocity [20, 21]. At each iteration, the adjoint gradient is evaluated from the interaction between two electromagnetic field distributions within the design region: the forward field excited by the incident OAM source and the adjoint field generated by the corresponding adjoint source. The optimization begins with a continuous material-density representation, initialized with a chiral-like shape to guide the search toward promising regions of the design space. In this grayscale formulation, the local permittivity is continuously varied between those of air and $Si_3N_4$, enabling gradient-based exploration using the Adam optimizer [22]. To progressively enforce a physically realizable binary structure, a spatial binarization constraint is introduced during the later stages of optimization. Specifically, a projection filter with a gradually increasing steepness parameter ($\beta$) is employed to drive the material distribution toward the two permittivity extremes.

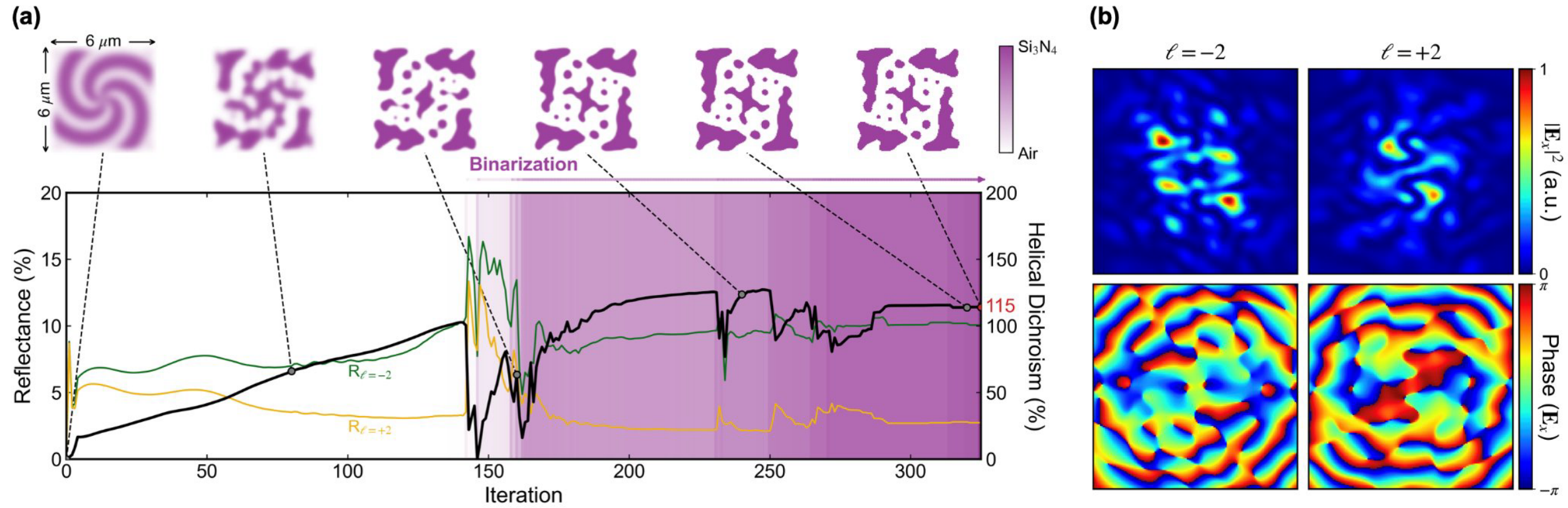

**Figure 1.** Evolution of the inverse-designed chiral metasurface and its chiroptical response. (a) Optimization trajectory showing the structural evolution from a continuous material distribution to the final binary layout (top) and the corresponding reflectance and helical dichroism (HD) performance (bottom). The background shading indicates the imposed binarization level. The final structure achieves an HD response of around 115%. (b) Simulated near-field electric-field intensity ($\mathbf{E}_x$) and phase distributions of the optimized structure under illumination with OAM modes of ($\ell = -2$) (left) and ($\ell = +2$) (right).

As presented in Figure 1(a), the optimization trajectory shows the progressive geometry updates driven by the gradient-ascent algorithm, while the background shading indicates the binarization level. Increasing the binary constraint temporarily perturbs the scattering phase profile, causing transient reductions in the HD metric, particularly for larger binarization increments. The optimization ultimately converges to a fully binary layout, as shown in the rightmost structural snapshot of Fig. 1(a), yielding a strong HD response. Fig. 1(b) further illustrates the origin of this asymmetric response by presenting the near-field electric-field intensity and phase distributions calculated using FDTD. The results reveal markedly different responses for the two opposing OAM topological charges, with stronger reflected-field excitation for the $\ell = -2$ mode and suppressed reflection for the $\ell = +2$ mode.

**Fabrication and Optical Characterization**

To experimentally validate the inverse-designed structure, the chiral metasurfaces were fabricated on quartz substrates using electron-beam lithography (EBL) (see the Methods section for details). The structures were arranged in an array with sufficient spacing of over 20 $\mu$m between neighboring elements to minimize near-field coupling while providing processing redundancy and facilitating optical characterization. Figure 2(a) shows a large-area scanning electron microscopy (SEM) image of the fabricated array, with the corresponding identifier indicated below the patterns. A magnified SEM image in Fig. 2(b) confirms the successful fabrication and reproduction of the designed chiral geometry.

Following structural characterization, wide-field optical images of the fabricated metasurfaces were acquired using a color CMOS camera under white-light illumination, as shown in Fig. 2(c) and (d). To experimentally verify the spatial alignment of the structured-light excitation, an LG beam carrying an OAM state of $\ell = -2$ was focused onto the sample. The beam was first positioned on the bare substrate adjacent to the metasurface array, as indicated by the black outline in Fig. 2(c), to verify its focal position and profile. With the white-light illumination turned off, the corresponding inset provides a magnified view of the focused beam profile. The sample stage was then moved laterally to position the focused LG beam onto a selected chiral metasurface, as indicated by the black outline in Fig. 2(d). The corresponding inset, acquired without the white-light background, clearly shows the spatial overlap between the focused beam and the targeted structure, confirming the alignment prior to spectral measurements.

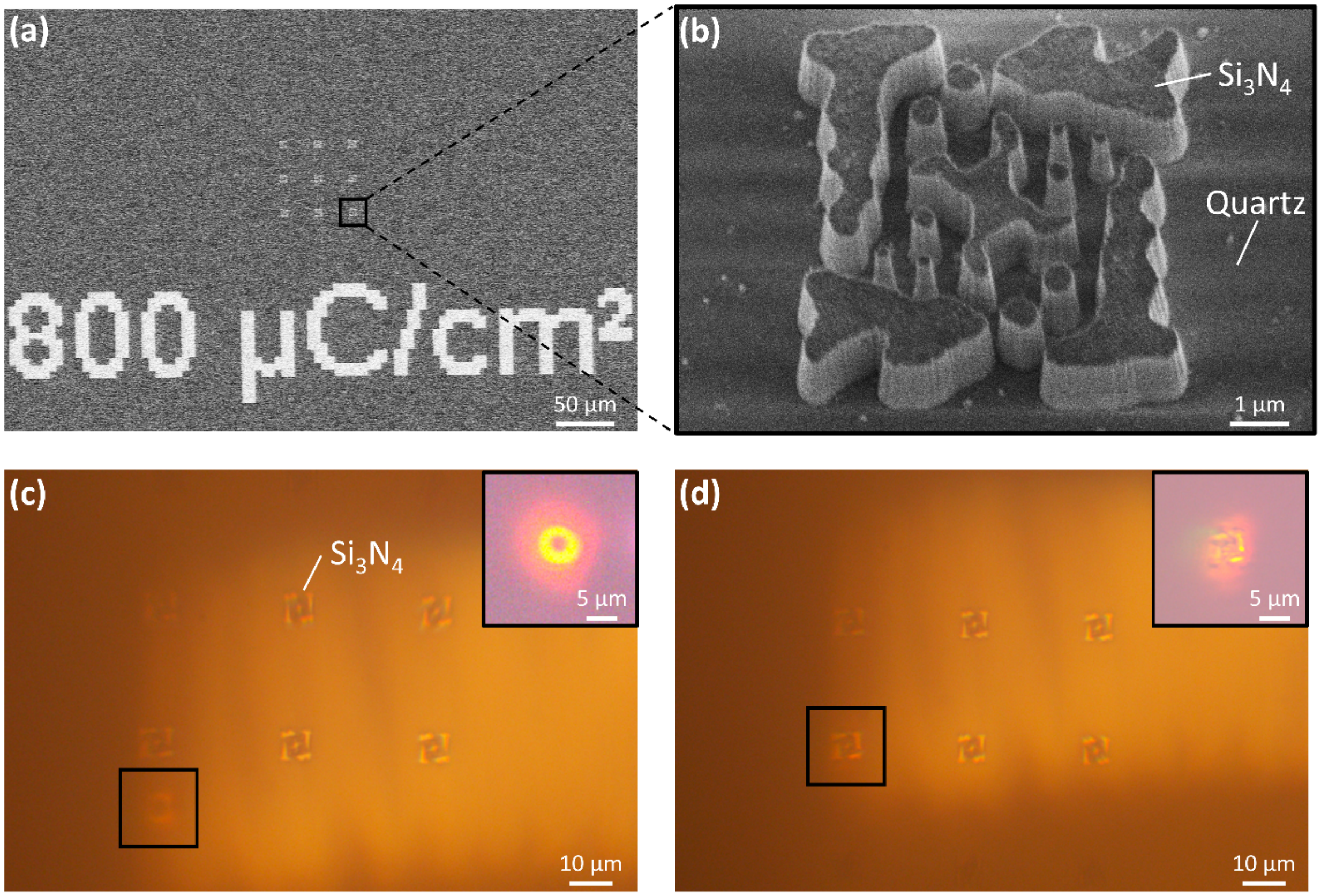

**Figure 2.** Optical characterization of the fabricated chiral metasurfaces. (a) Scanning electron microscopy (SEM) image of the metasurface arrays, with the corresponding array identifiers indicated below. (b) Magnified SEM image of the fabricated chiral structure. (c) Optical microscopy image of the metasurface array acquired using a color CMOS camera under white-light illumination. The inset shows the focal profile of the incident LG beam with an OAM state of $\ell = -2$, focused onto the bare substrate adjacent to the metasurface array for reference. (d) Optical microscopy image showing the spatial alignment of the focused LG beam with a selected chiral metasurface. The inset shows a magnified view of the beam-structure overlap with the white-light illumination turned off.

**Experimental Observation of Strong Helical Dichroism**

To experimentally characterize the chiroptical response of the fabricated inverse-designed structure under structured-light excitation, a custom-built optical setup was employed, as illustrated in Figure 3(a) and further detailed in the Methods section. An 800 nm excitation beam from a supercontinuum laser was first prepared with a half-wave plate (HWP) and a linear polarizer (LP) to obtain a well-defined linear polarization before illuminating the spatial light modulator (SLM). The SLM was programmed to generate either of the two LG beams carrying opposite OAM states of $\ell = -2$ and $\ell = +2$, as shown by the calculated phase profiles in Fig. 3(b) and (c), respectively. To experimentally generate and spatially isolate the desired LG mode, the corresponding phase profiles were encoded onto the SLM using computer-generated blazed-grating holograms, shown in Fig. 3(d) and (e). The hologram consisted of the LG phase profile superimposed with a tilted phase ramp, which separates the desired first-order diffracted beam from the zeroth-order reflection. The detailed hologram design is described in the Methods section.

The generated first-order LG beam was subsequently directed toward the sample and focused onto a single chiral structure using an objective lens (OL). An integrated white-light imaging path, incorporating a halogen source, a CMOS camera, and a removable beamsplitter, was used to facilitate spatial alignment of the incident beam with the target structure. Once the beam was accurately aligned, the removable beamsplitter was detached to minimize optical losses in the reflected signal during spectral measurements. The reflected optical signal was collected by the same OL, coupled into a multimode fiber and recorded using a spectrometer. Reflectance spectra were obtained for both $\ell = -2$ and $\ell = +2$ excitation, enabling direct comparison of the structure's chiroptical response to opposite OAM states.

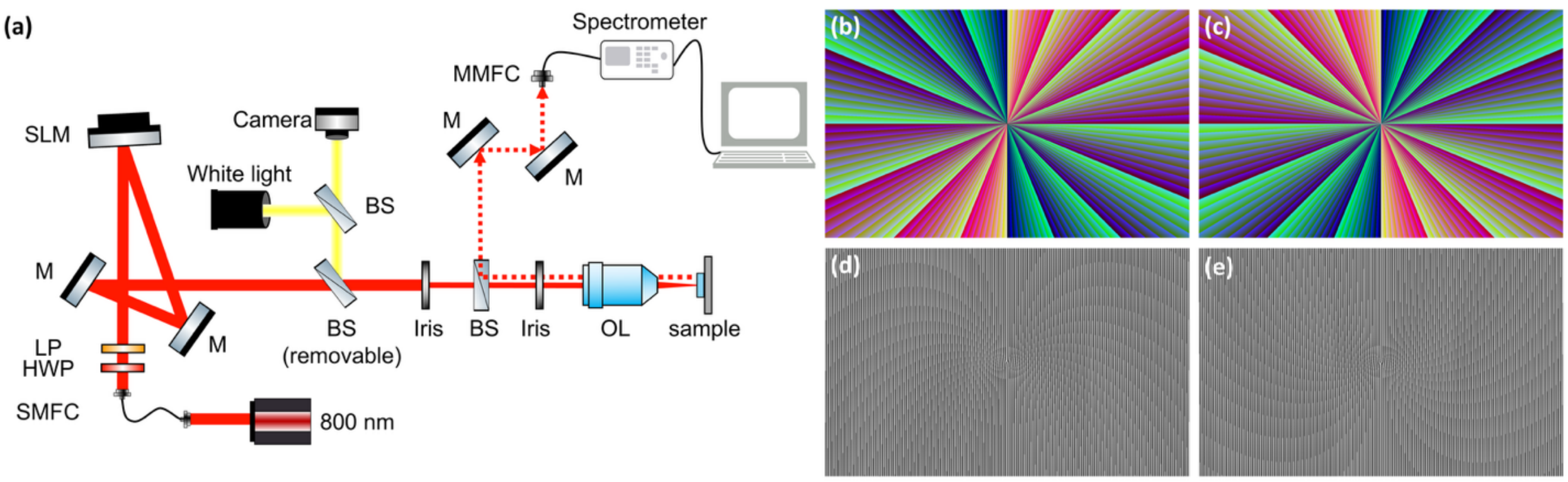

**Figure 3.** Schematic of the experimental setup and spatial light modulator (SLM) phase patterns for Laguerre-Gaussian (LG) beam generation. (a) Schematic of the custom-built optical setup for OAM-carrying LG beam generation and reflectance measurements. The SMFC, HWP, LP, M, BS, OL, and MMFC denote the single-mode fiber collimator, half-wave plate, linear polarizer, mirrors, beamsplitter, objective lens, and multimode fiber collimator, respectively. An integrated white-light imaging path with a removable beamsplitter enables spatial alignment of the incident beam onto the sample. (b) Calculated phase profile for generating an LG beam with an OAM state of $\ell = -2$. (c) Calculated phase profile for generating an LG beam with an OAM state of $\ell = +2$. (d) Corresponding computer-generated blazed-grating hologram for the $\ell = -2$ LG beam. (e) Corresponding computer-generated blazed-grating hologram for the $\ell = +2$ LG beam.

To maximize the signal-to-noise ratio and minimize contributions from ambient light and stray illumination, all room and halogen light sources were switched off during data acquisition. Background spectra were also recorded and subtracted from the measured spectra to account for detector background and stray-light contributions. As a control experiment, a geometrically symmetric feature of an adjacent EBL identifier was measured as an achiral reference under the same experimental conditions. As shown in Figure 4(a), the reflectance spectra under $\ell = -2$ and $\ell = +2$ LG illumination were nearly identical. The corresponding HD spectrum in Fig. 4(b) remains near zero over most of the measured spectral range, with small fluctuations near the design wavelength. The small response indicates that the observed spectral variations in the achiral reference do not produce an apparent differential response between the two OAM states, providing a suitable control for subsequent characterization of the chiral metasurface. Following this measurement, the excitation beam was positioned on a single isolated chiral metasurface unit cell. As shown in the inset of Fig. 2(d), the focused LG beam was spatially centered on the target structure. Under this excitation condition, the measured reflectance spectra in Fig. 4(c) exhibit a clear dependence on the OAM state, with substantially higher reflectance for the $\ell = -2$ illumination near the design wavelength. Correspondingly, the HD spectrum in Fig. 4(d) exhibits a distinct peak, reaching 52.6% at 800 nm and a maximum value of 61.7% at 801.6 nm. These measurements demonstrate a pronounced OAM-dependent helical dichroic response from a single planar chiral metasurface.

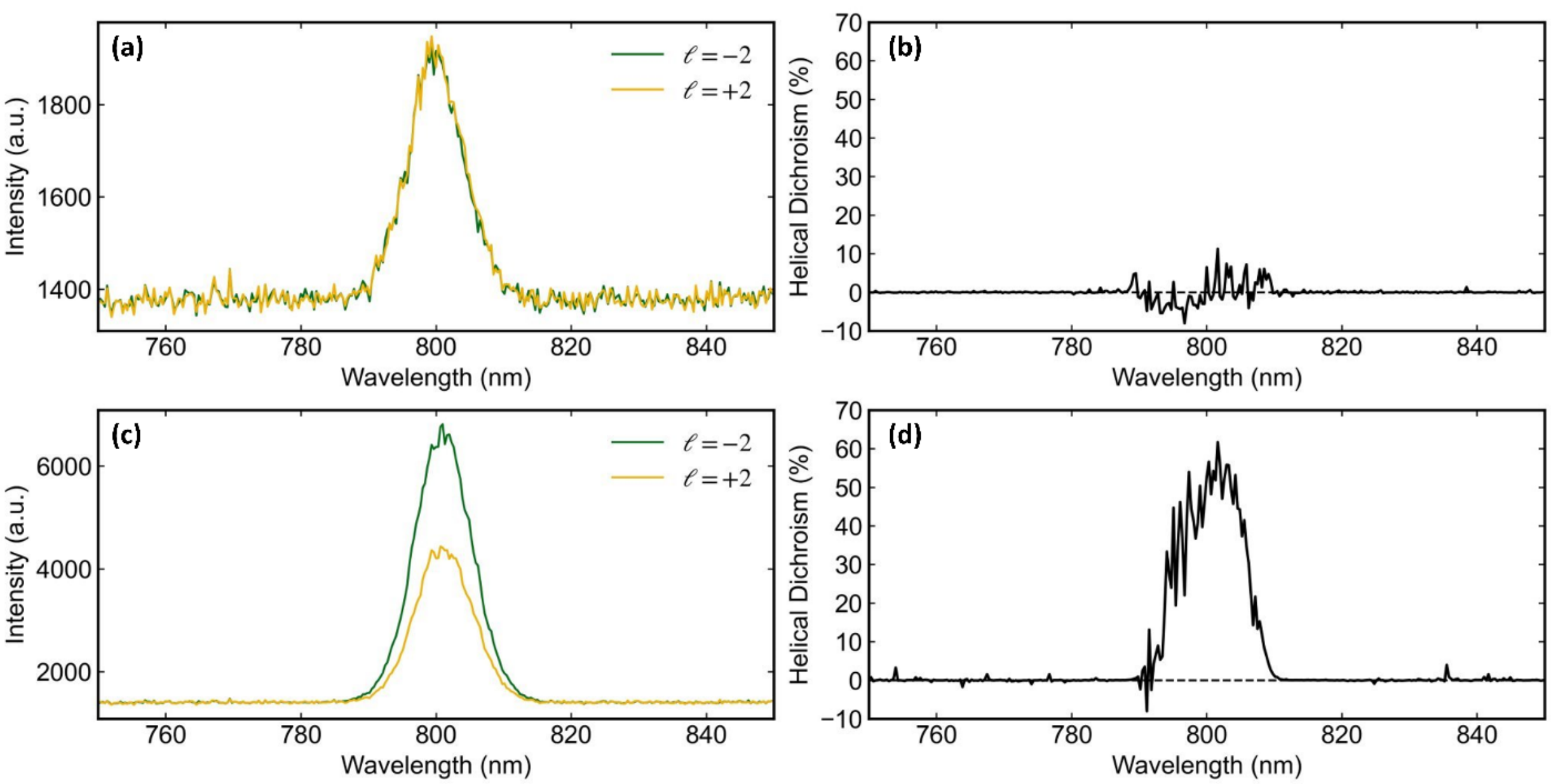

**Figure 4.** Experimental measurement of the HD response. (a) Measured reflectance spectra of a geometrically symmetric feature of an adjacent EBL identifier used as an achiral reference, under the illumination of LG beams carrying OAM states of $\ell = -2$ (green) and $\ell = +2$ (yellow). (b) The corresponding HD spectrum derived from the reflectance measurements in (a), showing a marginal chiroptical response from the achiral reference. (c) Measured reflectance spectra of a single fabricated chiral metasurface under the same LG beam illuminations. (d) Corresponding HD spectrum of the chiral metasurface, exhibiting an HD of 52.6% at 800 nm and reaching a maximum of 61.7% at 801.6 nm.

In particular, the measured HD is lower than the maximum value of approximately 115% obtained in the simulation. This discrepancy may arise in part from geometric differences between the simulated and fabricated structures. As observed in the SEM image in Fig. 2(b), features such as sidewall tapering, edge roughness, and corner rounding may alter the local electromagnetic field distribution and resonance conditions relative to the simulated structure. Despite the discrepancy between the experimental and simulated results, the measured HD of up to 62% demonstrates the potential of inverse design to enhance OAM-dependent chiroptical interactions in compact planar dielectric metasurfaces.

**CONCLUSION**

In conclusion, we demonstrated an inverse-designed chiral metasurface capable of producing enhanced helical dichroism under orbital-angular-momentum-carrying illumination. By incorporating the spatial characteristics of Laguerre-Gaussian beams into the inverse-design framework, the chiral geometry was optimized to achieve a simulated HD of about 115% at 800 nm for opposite OAM states. The optimized structures were subsequently fabricated in $Si_3N_4$ and experimentally characterized using a custom-built optical system. An achiral reference structure exhibited a minor HD response, while a single fabricated chiral metasurface produced a maximum HD of nearly 62% near 800 nm, confirming a substantial OAM-dependent chiroptical response at the single-structure level. The experimental results demonstrate the feasibility of using inverse-designed planar chiral metasurfaces to enhance light-matter interactions with structured optical fields. This work provides a practical pathway toward ultra-compact, on-chip platforms for chiroptical sensing and OAM-resolved imaging with CMOS-integrated photonic systems.

## METHODS

### Fabrication

The chiral metasurfaces were fabricated on quartz substrates coated with a 1.2 $\mu$m-thick $Si_3N_4$ layer deposited by plasma-enhanced chemical vapor deposition (Oxford Instruments PLASMALAB 100 PECVD). To realize the designed structures with an aspect ratio of approximately 4, a bilayer polymethyl methacrylate (PMMA) electron-beam resist was prepared on the $Si_3N_4$ surface. EBL (Raith EBPG5000plusES) was then used to define the chiral structures together with the identifiers, followed by cold development in ethanol at 4 °C for 45 s, after which the sample was thoroughly dried with nitrogen. A nickel (Ni) layer was subsequently deposited by electron-beam evaporation (Intlvac Nanochrome II) to serve as the etch mask [23]. Lift-off was performed and then the structures were etched using inductively coupled plasma reactive-ion etching (Oxford Instruments Plasmalab 100 - ICP380), transferring the designed pattern into the $Si_3N_4$ layer [24]. Finally, the residual Ni etch mask was removed using a Piranha solution.

### Experimental Setup

A supercontinuum laser (SuperK FIANIUM, NKT Photonics) equipped with a SuperK VARIA wavelength filter was used as the excitation source and spectrally filtered to a center wavelength of 800 nm with a bandwidth of 10 nm. The beam was outcoupled through a single-mode fiber collimator and polarized using a half-wave plate (Thorlabs AHWP05M-600) and linear polarizer before illuminating the liquid-crystal-on-silicon spatial light modulator (Santec SLM-200). LG beams with OAM states were generated using computer-generated holograms displayed on the SLM [25, 26]. The holograms were constructed by superimposing the calculated LG phase profile with a tilted plane-wave phase to form a blazed grating. A diffraction angle of 0.5° was introduced to spatially separate the desired first-order diffracted beam from the zeroth-order component. An incident beam diameter of approximately $D = 1.8$ mm before the 50× Mitutoyo M Plan Apo objective (NA = 0.55, $f = 4.0$ mm, WD = 13 mm) was considered. Because the beam diameter was smaller than the objective aperture, the objective was effectively underfilled, giving an effective $\mathrm{NA_{eff}}$ of about 0.225, calculated as $\mathrm{NA_{eff}} \approx D/(2f)$. The corresponding beam waist at 800 nm was estimated as $w_0 = \lambda/(\pi \mathrm{NA_{eff}}) \approx 1.13\ \mu\mathrm{m}$. The reflected signal was collected through the same objective and coupled into a multimode fiber collimator connected to a spectrometer (Ocean Optics HR6). The measured spectra were background-subtracted and normalized to the corresponding Au reference spectrum to obtain the normalized reflectance for each OAM state. For spatial alignment, a white-light imaging path consisting of a halogen source and CMOS camera (Thorlabs CS165CU/M) was introduced through a removable beamsplitter. The beamsplitter was mounted on a magnetic indexing base to enable reproducible positioning and was removed after alignment to minimize attenuation of the reflected signal.

## AUTHOR CONTRIBUTIONS

C.-C.P. and S.K. conceived the idea. C.-C.P. contributed to the inverse-design methodology, performed the fabrication and optical measurements, analyzed the data, prepared the figures, and wrote the manuscript. Y.-H.C. contributed to the inverse-design methodology, simulation validation, data analysis, figure preparation, and manuscript writing. M.B. contributed to the inverse-design methodology and contributed to writing the manuscript. D.S. contributed to fabrication process development and optimization. D.F. contributed to the design and troubleshooting of the optical setup. H.C. contributed to project supervision and manuscript review. S.K. contributed to conceptualization, project supervision, and manuscript writing and review. R.R.U. supervised the project and contributed to manuscript writing and review. All authors discussed the results and commented on the manuscript.

## ACKNOWLEDGMENT

Computational work was performed using the high-performance computing (HPC) resources of the University of Melbourne (Project ID: punim2399). This work was performed in part at the Melbourne Centre for Nanofabrication (MCN) in the Victorian Node of the Australian National Fabrication Facility (ANFF).